\documentclass[12pt,aps,prb,preprint]{revtex4}   % style for Physical Review B and AJP are similar

\usepackage{amsmath}    % need for subequations
\usepackage{graphicx}   % for figures

\begin{document}

\title{Stochastic motion of test particle implies that G varies with time}
\author{Davood Momeni}
 %\altaffiliation[Also at ]{home.}  %  optional
 %\affiliation{Department of physics,Faculty of basic
%sciences,Tarbiat Moa'llem university,Tehran,IRAN}
 \email{d.momeni@yahoo.com}   %optional
%\author{Jan Tobochnik}
%\affiliation{Kalamazoo College, Department of Physics, Kalamazoo, MI
%49007}
\date{\today}

\begin{abstract}
The aim of this letter is to propose a new description to the time
varying gravitational constant problem, which naturally implements
the Dirac's large numbers hypothesis in a new proposed holographic
scenario for the origin of gravity as an entropic force. We survey
the effect of the Stochastic motion of the test particle in
Verlinde's scenario for gravity\cite{Verlinde}. Firstly we show that
we must get the equipartition values for $t\rightarrow\infty$ which
leads to the usual Newtonian gravitational constant. Secondly,the
stochastic (Brownian) essence of the motion of the test particle,
modifies the Newton's 2'nd law. The direct result is that the
Newtonian constant has been time dependence in resemblance as
\cite{Running}.
\end{abstract} \maketitle

\section{General remarks about the Dirac Large Number
Hypothesis(LNH)}
 During our survey of various physical laws of
nature, we come across a number of constants entangled with those
laws. Historically it was Weyl \cite{Weyl1917,Weyl1919} who
initiated the idea of large numbers. But, it was Dirac who
discovered an apparently unseen thread joining up those physical
constants by a simple yet interesting law, viz., \emph{Law of Large
Numbers}. Using that law, Dirac arrived at his \emph{Large Number
Hypothesis} which has profound influence on the world of physics. In
fact, a plethora of works have been done, both at theoretical and
observational level with LNH as their starting point. Some works are
centered around modification of Einstein's gravitational theory and
related equations for adopting the idea of G
variation\cite{Dirac1973},\cite{Canuto}. Another class of works
counter arguments  against LNH for justifying and refuting that
hypothesis characterize the second category of works in favor of it
\cite{Bousso} and also Testing the validity of it \cite{Blake}. Even
there is at least a formal  link between LNH and cosmological
constant term $\Lambda$ \cite{Peebles}. The Weyl's large number is
root square of Eddington's one \cite{Eddington} that was supported
by observational data \cite{Stewart}. Indeed Stewart showed  that
the ratio of the radius of the universe and electron was only two
orders of magnitude smaller $10^{40}$ than that of Weyl's number.
Eddington's magic number is  $N = 1.7507\times10^{85}$ and Weyl's
number is $\sqrt{N}$. There are some other large numbers , Jordan's
number \cite{Jordan}and recently the Shemi-zadeh's number
\cite{Shemi-zadeh}. There is a wide class of acclaim about the
relation between LNH and other concepts of theoretical physics. For
example as was claimed by G$\ddot{o}$rnitz , it may be exist a
\emph{close connection between the Bekenstein-Hawking entropy and
Weizs$\ddot{a}$ckers ur theory} \cite{Thomas},\cite{Weizsacker}.
There is  almost  a new and updated review which contains some of
these aclaims and theories \cite{Ray}.
\subsection{ Formulation of Dirac Large Numbers Hypothesis and time running of the $G$}
There are three dimensionless numbers in the nature which can be
constructed from the atomic and cosmological datas:\\
1-The ratio of the electric to the gravitational force between an
electron and a proton $7\times10^{39}$\\
2-The age of the Universe t, expressed in terms of a unit of time
provided by atomic constants $\frac{e^{2}}{m_{e}c^{3}}$\\
and finally \\
3-\emph{ The mass of that part of the Universe that is receding from
us with a velocity $v<c/2$
expressed in units of the proton mass of the order $10^{78}$}\\
Dirac large number
hypothesis in it's orthodoxies form Germaned by himself says that\\
"..\emph{these numbers are related by equations in which the
coefficients are close to unity}". Since the number in (2) varies
with the age of the Universe, the L.N.H. requires that the other
numbers must also vary, namely
\begin{eqnarray}
\frac{e^{2}}{Gm_{e}m_{p}}\propto t
\end{eqnarray}
or
\begin{eqnarray}
N\propto t^{2}\\
G\propto t^{-1}
\end{eqnarray}
There are two interpretations for the above relation both discussed
by Dirac and the only one which was acceptable by himself as
\emph{One can reconcile the relation (2) with conservation of mass
by assuming that the velocity of recession of a galaxy is
continually decreasing, so that more and more galaxies are
continually appearing
with velocity of recession $<c/2$. }\\
This is the picture which was adopted in his first paper on the
subject \cite{Dirac,Dirac1973}. There are a serious problem between
(3) and GR: the Einstein's theory requires G to be constant. As was
noted by Dirac's theory this inconsistency might be solved
if\emph{".. we assume that the Einstein theory is valid in a
different system of units from those provided by the atomic
constants."} Consequence of Dirac's LNH is the coexistence of a
Variable G Cosmology.
\section{Entropic approach to the origin of gravity}
In \cite{Verlinde} it is postulated that the change of entropy,
related to the entropy that is saved on the holographic screen,
satisfies the following relations
\begin{equation}
\Delta S = 2 \pi k_{B}, \quad \Delta x = \hbar/m c.
\end{equation}
The coefficient $2 \pi$ is stipulated by matching the correct
expression for the force $F$
\begin{equation}
F \Delta x = T \Delta S.
\end{equation}
The temperature is associated with the acceleration through the
famous Unruh formula \cite{Unruh}
\begin{equation}
k_{B} T = \frac{\hbar a}{2 \pi c}.
\end{equation}
Implying the homogeneous distribution of information on the
holographic screen,  for a particle approaching the screen one can
write
\begin{equation}
m c^2 = \frac{1}{2} n k_{B} T,
\end{equation}
where $n$ is the number of bits. Together with the Unruh formula
this gives
\begin{equation}
\frac{\Delta S}{n} = k_{B} \frac{a \Delta x}{2c^2}.
\end{equation}
In the general relativistic context one starts from a generalized
form of the Newtonian potential
\begin{equation}
\phi = \frac{1}{2} \log (-g^{\alpha \beta }\xi_{\alpha}
\xi_{\beta}),
\end{equation}
where $e^{\phi}$ is the red-shift factor that is supposed to be
equal to unity at the infinity ($\phi =0$ at $r= \infty$), if the
space-time is asymptotically flat. The background metric is supposed
to be some static solution which admits a global time-like Killing
vector $\xi_{\alpha}$.

The acceleration is defined by the formula
\begin{equation}
a^{\alpha} = - g^{\alpha \beta} \Delta_{\beta} \phi,
\end{equation}
and the Unruh-Verlinde temperature on the screen is given by the
formula
\begin{equation}\label{T}
T = \frac{\hbar}{2 \pi} e^{\phi} n^{\alpha} \Delta_{\alpha} \phi,
\end{equation}
where $n_{\alpha}$ is a unit vector, that is normal to the
holographic screen and the Killing time-like vector $\xi_{\beta}$.
In this approach it is supposed that the motion of the test particle
is classical and no friction like force exists in the spacetime.In
the next section we generalized the classical motion of test
particle to a stochastic's one, and we will show that the no-long
time limit of the motion leads to a time dependent gravitational
constant $G$. This step is obtained by replacing the usual
equipartition law with another time dependent formula which is valid
not only in classical regime $t\rightarrow\infty$ but for other
finite time interval.
\section{On the stochastic motion of test particle}
The Roman Lucretius's scientific poem "On the Nature of Things"
\cite{Lucretius} has a remarkable description of Brownian motion of
dust particles. He uses this as a proof of the existence of atoms:

"\textrm{Observe what happens when sunbeams are admitted into a
building and shed light on its shadowy places. You will see a
multitude of tiny particles mingling in a multitude of ways... their
dancing is an actual indication
 of underlying movements of matter that are hidden from our sight...
 It originates with the atoms which move of themselves .
 Then those small compound bodies that are least removed from the impetus
of the atoms are set in motion by the impact of their invisible
blows and in turn cannon against slightly larger bodies. So the
movement mounts up from the atoms and gradually emerges to the level
of our senses, so that those bodies are in motion that we see in
sunbeams, moved by blows that remain invisible.}"\\
 Although the mingling motion of dust particles is caused largely by air currents,
the glittering, tumbling motion of small dust particles is, indeed,
caused chiefly by true Brownian dynamics. The first person to
describe the mathematics behind Brownian motion was Thiele
\cite{Thiele} in a paper on the method of least squares.  However,
it was  Einstein \cite{Einstein}and  Smoluchowski
\cite{Smoluchowski} who independently brought the solution of the
problem to the attention of physicists, and presented it as a way to
indirectly confirm the existence of atoms and molecules.
Specifically, Einstein predicted that Brownian motion of a particle
in a fluid at a thermodynamic temperature T  is characterized by a
diffusion coefficient
\begin{eqnarray}
D = \frac{k_{B }T}{b}
\end{eqnarray}

where $k_{B }$ is Boltzmann's constant and b is the linear drag
coefficient on the particle (in the Stokes/low-Reynolds regime
applicable for small particles). As a consequence, the root mean
square displacement in any direction after a time t is
\begin{eqnarray}
\overline{s^{2}}=2Dt=\frac{2 k_{B}T}{f}t
\end{eqnarray}
At first the predictions of Einstein's formula were seemingly
refuted by a series of experiments, which gave displacements of the
particles as 4 to 6 times the predicted value. But Einstein's
predictions were finally confirmed in a series of experiments
carried out by Chaidesaigues \cite{Chaidesaigues}and Perrin
\cite{Perrin} . The confirmation of Einstein's theory constituted
empirical progress for the kinetic theory of heat. In essence,
Einstein showed that the motion can be predicted directly from the
kinetic model of thermal equilibrium. The importance of the theory
lay in the fact that it confirmed the kinetic theory's account of
the second law of thermodynamics as being an essentially statistical
law. For more physical examples specially for applications of
stochastic problems in physics we refer the reader to the classical
review of Chandrasekhar which is the best one even after near 70
years\cite{Chandrasekhar}.
\subsection{Modeling using differential equations}
In mathematics, Brownian motion is described by the Wiener process;
a continuous-time stochastic process named in honor of Norbert
Wiener\cite{Wiener}. The Wiener process $W_{t}$ is characterized by
this fact:
\begin{eqnarray}
 W_{0} = 0
 \end{eqnarray}
 $W_{t}$ is almost surely continuous
Also $W_{t}$ has independent increments. An alternative
characterization of the Wiener process is the so-called Levy
characterization that says that the Wiener process is an almost
surely continuous martingale with $W_{0}$ and quadratic variation
$[W_{t},W_{t}] = t$.

A third characterization is that the Wiener process has a spectral
representation as a sine series whose coefficients are independent
$\mathcal{N}(0, 1)$ random variables. This representation can be
obtained using the Karhunen–Loeve theorem.

The Wiener process can be constructed as the scaling limit of a
random walk, or other discrete-time stochastic processes with
stationary independent increments. This is known as Donsker's
theorem. Like the random walk, the Wiener process is recurrent in
one or two dimensions (meaning that it returns almost surely to any
fixed neighborhood of the origin infinitely often) whereas it is not
recurrent in dimensions three and higher. Unlike the random walk, it
is scale invariant.

The time evolution of the position of the Brownian particle itself
can be described approximately by a Langevin equation, an equation
which involves a random force field representing the effect of the
thermal fluctuations of the solvent on the Brownian particle. On
long timescales, the mathematical Brownian motion is well described
by a Langevin equation. On small timescales, inertial effects are
prevalent in the Langevin equation. However the mathematical
Brownian motion is exempt of such inertial effects. Note that
inertial effects have to be considered in the Langevin equation,
otherwise the equation becomes singular, so that simply removing the
inertia term from this equation would not yield an exact
description, but rather a singular behavior in which the particle
doesn't move at all.

\subsection{Ornstein's approach to the Brownian motion}
Following the Ornstein and Uhlenbeck method \cite{Ornstein} for
surveying the motion of the Brownian particle,we know that such a
particle obeys from the famous Einstein-Langevin
equation\cite{Einstein}:
\begin{eqnarray}
\frac{d u}{dt}=-\beta u+w(t)
\end{eqnarray}
Here $u(t)$ is the velocity of the particle.The influence of the
surrounding medium is split into two  distinct parts:\\
(1)A friction part $-\beta u$\\
(2)A fluctuating part $w(t)$.\\
 It must be understanding as a stochastic differential equation and not a
commonplace one. The mean(average) is taken over an ensemble
 of particles which have started at $t=0$ with the same    velocity $u_{0}$ as the initial velocity at $t=0$.
The force(per unit mass) of the particle is restricted such that it
is a random distributed function of time as it's average vanishes
and also it is momentous only for two neighboring correlation for
small time's intervals. The interaction of the particle with the
medium creates from a dissipative velocity dependence term $-\beta
u$ and a Random force $w(t)$. The first method to solve the problem
is by calculating all the mean values $\overline{u^{k}}$ for given
$u_{0}$. As has first been shown by Ornstein \cite{Ornstein}
 for $\overline{u}$ and $\overline{u^{2}}$, this is possible by
 integrating the equation of motion (15) . Of course, the next
 assumptions hold for the fluctuating acceleration $w(t)$:
 \begin{eqnarray}
\overline{ w(t)}^{u_{0}}=0\\
\overline{ w(t_{1}) w(t_{2})}^{u_{0}}=\phi_{1}(t_{1}-t_{2})
\end{eqnarray}
where $\phi_{1}(x)$ is a function with a very sharp maximum at
$x=0$. More generally, when $t_{1},t_{2},...t_{n+1}$ are all lying
very near each other, we assume:
\begin{eqnarray}
\overline{ \prod_{i=1}^{n+1}
w(t_{i})}^{u_{0}}=\phi_{n}(r,\theta_{1},\theta_{2},...,\theta_{n-1})
\end{eqnarray}
where r is the distance perpendicular to the line
$t_{1}=t_{2}=...=t_{n+1}$ in the (n+1) dimensional
$(t_{1},t_{2},...,t_{n+1})$ space, and
$(\theta_{1},\theta_{2},...,\theta_{n-1})$ are (n-1) angels to
determine the position of $r$ in the subspace perpendicular to this
line. The function $\phi_{n}$ has again a very sharp maximum for
$r=0$.

In brief
\begin{eqnarray}
\overline{\int ^{\infty}_{-\infty}w(t)dt}=0\\
\int^{\infty}_{-\infty}\overline{w(\xi)w(\xi+\psi)}d\xi=\theta
\end{eqnarray}
We know that the distribution function for such particles must be
Gaussian with mean and variance
\begin{eqnarray}
\overline{u(t)}=u_{0}e^{-\beta t}\\
\overline{u(t)^2}=u_{0}^2 e^{-2\beta t}+\frac{(1-e^{-2\beta
t})}{2\beta}\theta
\end{eqnarray}
The long time limit of the distribution function of particles must
be Maxwellian with temperature $T$.Thus we obtain the following
alterative form for equipartition theorem
\begin{eqnarray}
\overline{\frac{1}{2}mu(t)^2}=\frac{1}{2}m(u_{0}^2e^{-2\beta
t}+\frac{k_{B}T}{m}(1-e^{-2\beta t}))
\end{eqnarray}
Which it has the common form only for long times and differs very
stranger for finite times chiefly for short times after beginning
the motion.

\section{Brownian correction to the Newton's gravity via Verlinde's
approach} In this section we replace (7) with (23) in section (2).
Following the Verlinde's nice idea\cite{Verlinde} about the gravity
as an entropic force and gravity as an emergent phenomena we know
that if a test particle  accedes neat to a holographic screen (in
Verlinde original proposal,a collection of equipotential surfaces in
spacetime)which has the mass $M$ and the test particle seances
himself in a bath with Unruh temperature\cite{Unruh}and by assuming
that the holographic screen has N bits $N=\frac{A}{l_{p}^2}$(The
horizon ,if we take the equipotential surface as the surface of the
black-hole)is a sphere with radius $R$ and by replacing the
alternative-Brownian analogous of the equipartition theorem instead
of the infinite time approximation we obtain the next expression for
gravitational acceleration
\begin{eqnarray}
a=\frac{G_{eff}M}{R^2}-\frac{2\pi
u_{0}^2}{\lambda_{c}}\frac{e^{-2\beta t}}{1-e^{-2\beta t}}
\end{eqnarray}
Where in it
\begin{eqnarray}
G_{eff}=\frac{G_{N}}{1-e^{-2\beta t}}
\end{eqnarray}
is going to be identified with Newton's constant ,thus only for long
times.The$G_{eff}$ must be understood as an effective time varying
Newton's constant. $\lambda_{c}$ is the Compton's wavelength of the
test particle. For small values of $\beta$ we have
\begin{eqnarray}
G_{eff}=\frac{G_{N}}{2\beta t}\propto t^{-1}
\end{eqnarray}
Comparing it with the  Dirac hypothesis about the large numbers
\cite{Dirac} is very surprising. Again it seems that there is a
delicate relation between running of the Newtonian constant, Dirac's
Large Numbers Hypothesis and Verlinde scenario for the
gravity\cite{Running} for by the Brownian motion hypothesis for test
particle.As i think that this running scheme must be related to the
quantum corrections of the Verlinde's
idea\cite{Quantum,Vancea,Nicolini}. Dirac interpreted this to mean
that G varies with time as , and thereby pointed to a cosmology that
seems '\emph{designer-made}' for a theory of quantum gravity.
According to General Relativity, however, G is constant, otherwise
the law of conserved energy is violated. Dirac met this difficulty
by introducing into the Einstein equations a gauge function $\beta$
that describes the structure of spacetime in terms of a ratio of
gravitational and electromagnetic units. He also provided
alternative scenarios for the continuous creation of matter, one of
the other significant issues in LNH are\\
1-'\emph{additive}' creation (new matter is created uniformly
throughout space) \\
2-'\emph{multiplicative}' creation (new matter is created where
there are already concentrations of mass).

 In above we observed that this
effective G arisen when we take the motion of the test particle as a
Stochastic one. Attending that at sufficiently long times
$t>>1/\beta$ the second term is negligible and we recover the usage
form of Newton's gravity.There is an unsolved problem about the
appearance of the dissipation term $\beta$ in this equation.In the
classical theory of the Brownian motion we assume that the medium
obeys from the Stoke's formula for a test particle in an emulsion
medium. We can attribute this property to the nature of the
gravitational attraction or other unlikely feature with an unknown
pedigree.

\section{Rate of Change of $G_{eff}$}
A volume of works has been centered around the act of calculating
the amount of variation of the gravitational constant.See for
example the references  \cite{Gaztanaga,Arzoumanian,Stairs}

From (25) we get,
\begin{eqnarray}
\frac{\dot{G_{eff}}}{G_{eff}}=-\frac{2\beta}{e^{2\beta  t}-1}
\end{eqnarray}
For small values of $\beta$,equation (27) tells us that the rate of
change of G is of the order of $t^{-1}$. Also, the hypothesis
demands that creation of matter occurs continuously in the universe.
This creation of matter can occur in two possible ways, viz.,
"\emph{additive creation}" and "\emph{multiplicative creation}".
According to "\emph{additive creation theory}", matter is created
through the entire space and hence in intergalactic space also. In
"\emph{multiplicative creation theory}", creation of matter occurs
only in those places where matter already exists and this creation
proceeds in proportion to the amount and type of atoms already
existing there. According to general relativity, G is constant and
hence we cannot readily consider G as a variable quantity in
Einstein equation. To overcome this difficulty, Dirac considered two
metrics. The equations of motion and classical mechanics are
governed by the Einstein metric which remains unaltered while the
other metric, known as atomic metric, includes atomic quantities and
the measurement of distances and times by laboratory apparatus
\cite{Faulkner}.The interval ds(A) separating two events as
determined by apparatus in atomic system of units (a.s.u.) will be
different from the interval ds(G) between the same two events as
measured in the gravitational system of units (g.s.u.). This implies
that equations written in g.s.u. and a.s.u. cannot be used at a time
until one of them is converted to the other system of units
\cite{Rogachev}. The velocity of light is unity for both metrics.
Considering the case of a planet orbiting the sun, Dirac
\cite{Dirac1974} showed that the relationship of Einstein and atomic
metric was different for additive and multiplicative creation
theory. In terms of the atomic distance scale, the solar system is
contracting for the additive creation model while it is expanding in
multiplicative creation.

 If we take the present age of the
Universe as 14 Gyr, then  the value of
$\frac{\dot{G_{eff}}}{G_{eff}}$ is of the order of $10^{-11}$ per
year which is supported by various theoretical and observational
results even in higher dimensional Dark Energy Investigation with
Variable $\Lambda$ and G in GR \cite{Utpal}. In context of a
formalism has been developped for discussing the symmetries of
Galilei-invariant classical and quantum mechanical systems
\cite{Galilei} associated with the nonrelativistic spacetime picture
\cite{spacetime}, in refrence \cite{horvathy} it was shown that  the
metric associated to a time-varying gravitational constant G(t) is
conformally related to the $G_{0}$ case if and only if G(t) changes
according to the prescription of Vinti \cite{Vinti}, whose
particular case is Dirac$'$s suggestion. Concerning the
observationally determined increase of the Astronomical Unit, more
recent estimates from processing of huge planetary data sets by
Pitjeva \cite{Pitjeva1,Pitjeva2} point towards a rate of the order
of $10^{-1} myr^{-1}$.  It may be noted that my result for the
secular variation of the terrestrial radial position on the line of
the apsides would agree with such a figure by either assuming a mass
loss by the Sun of just  $-9\times10^{-14} yr^{-1}$ or a decrease of
the Newtonian gravitational constant $\frac{\dot{G}}{G}\approx
-1\times 10^{-13}yr^{-1}$ . Such a value for the temporal varia-tion
of G is in agreement with recent upper limits from Lunar Laser
Ranging \cite{Muller}  $\frac{\dot{G}}{G}= (2\pm7)\times
10^{-13}yr^{-1}$. The main  main result of Pitjeva
\cite{Pitjeva2009} is $\frac{\dot{G}}{G} = (-5.9 ± 4.4)
\times10{-14}yr^{-1}$. Gaztanaga et al. \cite{Gaztanaga}, relying on
data provided by SN Ia \cite{Perlmutter,Riess} have shown that the
best upper bound of the variation of G at cosmological ranges is
given by
\begin{eqnarray}
-10^{-11}\leq \mid\frac{\dot{G}}{G} \mid \leq 0
\end{eqnarray}
where z, the red-shift, assumes the value nearly equal to 0.5.
Observation of spinning-down rate of pulsar PSR J2019+2425 provides
the result \cite{Arzoumanian,Stairs}
\begin{eqnarray}
\mid\frac{\dot{G}}{G} \mid\leq (1.4-3.2)\times10{-11}yr^{-1}
\end{eqnarray}

.

\section{Summary}
In this note we assumed that the test particle in the Verlinde's
scenario obeys from the Einstein-Langevin equation ,i.e. has a
Brownian motion. This assumption modified the equipartition theorem
for finite times.After inserting this modification in the entropic
expression for Newton 2'nd law,we observe that the usual formal
acceleration suddenly voids the stationary form and the
gravitational constant $G$ been time dependent. For  long times the
common gravity recovered but for small values of $\beta$, $G\propto
t^{-1}$.This is the repetition of the Dirac's hypothesis about large
numbers.Thus we can state that Dirac large number hypothesis is a
direct result from basic Holographic scenario of gravity with this
further assumption about the Stochastic nature of the test
particle's motion. Since the Brownian treatment is completely true
for all times ,and we must get the equipartition values for
$t\rightarrow\infty$ then it is putting that we can treat the test
particle in spacetime as a particle in a medium with the same
viscosity properties as emulsion.

\section{Acknowledgments}
We are indebted to P. A. Horv$a'$thy and E. V. Pitjeva  for their
interests and helps.

\end{document}